\documentclass{article}

\usepackage{arxiv}

\usepackage[utf8]{inputenc} 
\usepackage[T1]{fontenc}    
\usepackage{hyperref}       
\usepackage{url}            
\usepackage{booktabs}       
\usepackage{amsfonts}       
\usepackage{nicefrac}       
\usepackage{microtype}      
\usepackage[title]{appendix}
\usepackage{graphicx}
\usepackage[bf]{caption}

\usepackage{doi}
\usepackage{etoolbox}
\usepackage[natbibapa]{apacite}
\AtBeginDocument{}
\renewcommand{\APACrefnote}[1]{}

\newtoggle{bibdoi}
\newtoggle{biburl}
\makeatletter
\newsavebox{\bib@url}
\newsavebox{\bib@doi}

\undef{\APACrefURL}
\undef{\endAPACrefURL}
\undef{\APACrefDOI}
\undef{\endAPACrefDOI}

\newenvironment{APACrefURL}
  {\global\toggletrue{biburl}\lrbox\bib@url}
  {\endlrbox}

\newenvironment{APACrefDOI}
  {\global\toggletrue{bibdoi}\lrbox\bib@doi}
  {\endlrbox}

\newcommand{\printinfo}{
  \iftoggle{bibdoi}{\usebox{\bib@doi}}{\usebox{\bib@url}}
  \togglefalse{bibdoi}
}

\AtBeginEnvironment{thebibliography}{
\pretocmd{\PrintBackRefs}{%
  \iftoggle{bibdoi}
    {\iftoggle{biburl}{\unskip\unskip}{}\usebox{\bib@doi}}
    {\iftoggle{biburl}{Retrieved from \usebox{\bib@url}}}{}
  \togglefalse{bibdoi}\togglefalse{biburl}%
}{}{}}
\makeatother

\title{Do Human Mobility Network Analyses Produced from Different Location-based Data Sources Yield Similar Results across Scales?}

\date{} 					

\renewcommand{\headeright}{}
\renewcommand{\undertitle}{}
\renewcommand{\shorttitle}{}

\hypersetup{
pdftitle={Do Human Mobility Network Analyses Produced from Different Location-based Data Sources Yield Similar Results across Scales?},
pdfsubject={},
}

\begin{document}
\maketitle

\begin{center}
{\Large
Chia-Wei Hsu\textsuperscript{a,*},
Chenyue Liu\textsuperscript{a}, 
Kiet Minh Nguyen\textsuperscript{a}, 
Yu-Heng Chien\textsuperscript{a},
Ali Mostafavi\textsuperscript{a}
\par}

\bigskip
\textsuperscript{a} Urban Resilience.AI Lab, Zachry Department of Civil and Environmental Engineering,\\ Texas A\&M University, 199 Spence St., College Station, TX 77843\\
\vspace{6pt}
\textsuperscript{*} correseponding author, email: chawei0207@tamu.edu
\\
\end{center}
\bigskip
\begin{abstract}
The burgeoning availability of sensing technology and location-based data is driving the expansion of analysis of human mobility networks in science and engineering research, as well as in epidemic forecasting and mitigation, urban planning, traffic engineering, emergency response, and business development. However, studies employ datasets provided by different location-based data providers, and the extent to which the human mobility measures and results obtained from different datasets are comparable is not known. To address this gap, in this study, we examined three prominent location-based data sources—Spectus, X-Mode, and Veraset—to analyze human mobility networks across metropolitan areas at different scales: global, sub-structure, and microscopic. Dissimilar results were obtained from the three datasets, suggesting the sensitivity of network models and measures to datasets. This finding has important implications for building generalized theories of human mobility and urban dynamics based on different datasets. The findings also highlighted the need for ground-truthed human movement datasets to serve as the benchmark for testing the representativeness of human mobility datasets. Researchers and decision-makers across different fields of science and technology should recognize the sensitivity of human mobility results to dataset choice and develop procedures for ground-truthing the selected datasets in terms of representativeness of data points and transferability of results.
\end{abstract}



\section{Introduction}
\label{sec:Introduction}
Human mobility is the key to analyzing human activities and social interactions. Large datasets capturing anonymized human movement in urban area provide the tools to better understand complex phenomena by analyzing human mobility, such as spatial structures \citep{barthelemy_structure_2016,louail_uncovering_2015} and response to epidemics \citep{eubank_modelling_2004}. Over the past decade, studies have analyzed aggregated population movements to reveal universal laws and models that account for the statistical characteristics. Both temporal and spatial aspects of individual human trajectories can be revealed based on analysis of large-scale mobility data \citep{schlapfer_universal_2021,song_modelling_2010}. The analyses of human mobility networks have informed fields such as urban planning, traffic engineering, and pandemic mitigation. The number of studies examining human mobility networks has grown exponentially and is expected to increase further in the future. Example of studies utilizing human mobility data include recreational visitation estimation \citep{monz_using_2019}, accessibility inequality \citep{akhavan_accessibility_2019}, mobility patterns \citep{wang_extracting_2019}, emergency vehicle scheduling under catastrophic situations \citep{yan_mobiambulance_2019}, emergent structures in cities \citep{fan_fine-grained_2021}, and limitations of urban mobility network modelling \citep{hsu_limitations_2021}.

Datasets on which the models are based are a critical aspect of human mobility network analyses. Studies utilize different data sources, and the results reported for different phenomena may not be directly comparable. Given the significance of human mobility network analysis for of science, engineering, urban planning, and pandemic mitigation, it is essential to learn the extent to which the results obtained from models produced from different datasets would vary. For example, in the United States, several commercial third-party location-based data providers exist; however, no direct study has been conducted to evaluate the variation of human mobility network analysis results across different datasets. Such a comparison is critical to the juxtaposition and integration of results across different studies. To address this important gap, the study examined human network analysis results based on three major U.S.-based datasets, Spectus, X-Mode, and Veraset, across three scales, macroscopic network measures, motif properties, and microscopic mobility features, for multiple metropolitan areas in the United States.

\section{Background}
\label{sec:Background}
Advancements in sensing, mobile devices, and positioning technologies have made collection of passively generated datasets possible. The source of datasets include mobile phone call detail records (CDR) data \citep{jiang_activity-based_2017}, taxi/fleet GPS trajectories \citep{tang_uncovering_2015,yuan_discovering_2012}, Wi-Fi, and social media. These datasets provide new opportunities to understand human mobility patterns at a low cost and large scale \citep{xu_understanding_2015}. Examining complex patterns within large-scale, high-dimensional mobility data requires the use of advanced analysis techniques, usually based on network analysis and machine-learning methods. Applications of human mobility network analysis, based on similar characteristics, include user modeling, place modeling, and trajectory modeling \citep{toch_analyzing_2019}. Data containing massive spatial and temporal information can be further analyzed to study traffic distribution patterns \citep{tang_uncovering_2015}, distribution of functional regions in cities \citep{yuan_discovering_2012}, the spread of individuals’ activity space and its association with socio-economic status \citep{xu_understanding_2015}.

Since the outbreak of the COVID-19 pandemic in 2020, we have observed, more than ever, the significance of movement analytics and human mobility insights in fields including crisis mitigation and public health. Before the pandemic, the challenges for the researcher were data accessibility and the development of mobility analytic approaches. Access to detailed mobility data at fine granularity had raised concerns regarding ethics and privacy \citep{bertino_security_2008}, as well as representativeness and adequacy \citep{dodge_data_2021,lu_understanding_2017}. After the pandemic, a notable change occurred in the manner in which mobility data can be shared and accessed. In the United States., commercial companies, such as Meta, Spectus, SafeGraph, Google, Apple, and Descartes Labs, supported open data and data for good initiatives. Among those human mobility datasets, Spectus, X-Mode, and Veraset are the most representative ones and have been widely adopted by researchers and scientists. Researchers utilized these datasets and investigated human mobility change and social impacts due to the COVID-19 pandemic itself (emergence and severity) and related regulations and policies. Studies related to social impact delved into socio-economic gaps in mobility reduction \citep{fraiberger_uncovering_2020} and mobility change and association with social distancing \citep{bourassa_social_2020}. Studies related to regulations and policies investigated mobility changes and association with national lockdown \citep{pepe_covid-19_2020}, mobility and control measures on traffic safety \citep{zhang_effect_2021}, modeling human mobility trends under non-pharmaceutical interventions \citep{hu_big-data_2021}, the impact of restricting staff mobility between nursing homes \citep{jones_impact_2021}, and public and private responses to government actions \citep{gupta_tracking_2021}. Studies related to predicting the future state of the pandemic include early warnings for reopening of U.S. universities \citep{mehrab_high_2020}, evaluating proximity metrics in predicting the spread \citep{mehrab_evaluating_2021}, opportunities for local containment \citep{fan_neural_2021}, and spatiotemporal risk scores \citep{rambhatla_toward_2022}. 

Given the diversity of datasets used in human mobility studies, it is important to examine whether the conclusions and insights from different datasets are comparable and transferrable. Outputs of human mobility analyses could be sensitive to the source of the location-based dataset. For the science and practice around human mobility networks to advance, we should understand the extent to which the results produced from different datasets are similar or vary. To fill this gap and to provide evidence for comparing results reported across studies based on different datasets and to determine whether the results can be compared and generalized, this study examined the basic network properties generated from different datasets and their distribution on the global, substructure and microscopic level generated from different datasets.

This study is designed to verify whether analysis results and findings derived from different human mobility datasets can be compared and generalized to confidently allow formulation of global theories. Accordingly, we focus on addressing the following questions: (1) To what extent are the outputs of human mobility network analyses at different scales sensitive to different datasets? and (2) Which type of analysis at what scale is more sensitive to dataset type? Our hypotheses are: (1) The sensitivity of the observed features may increase as we analyze the human mobility network at a more granular level for all three datasets, (2) The overall distribution and fluctuation patterns of the observed features may be more similar between datasets that are collected with similar techniques. If the above hypotheses cannot be rejected, then we may conclude that the results and conclusions derived from different datasets and types of analyses cannot be directly generalized.

We selected three location-based datasets from three major providers in the United States: Spectus, X Mode and Veraset. These datasets are used extensively by researchers and are shown to representatively capture human movements for performing network analysis at the global, substructure, and microscopic levels. In this study, the datasets are aggregated daily to the census-tract level to construct human mobility networks and to obtain essential network features and characteristics for further analyses at various levels. At the macroscopic level, global features that characterize the entire human mobility network, including network size, average degree, average clustering coefficient, average shortest path length, assortativity coefficient, network modularity, network density, network diameter, and giant component size are calculated. At the substructure level, important feature categories are motif type distribution and motif features, such as travel distance and travel volume of each type of motif. At the microscopic level, we analyze the mobility features, such as average trip frequency, average travel distance, average travel time, and average radius of gyration.

\section{Data description and methods}
\label{sec:Data description and methods}
\subsection{Data collection}
The anonymized device-level location-based mobility data of February 2020 utilized throughout the following analyses were collected from three sources. We aggregated these data daily at the census-tract level then calculated trip frequencies between census tracts and constructed city-scale human mobility networks for calculating mobility metrics. We extracted data from February 2020 to test our hypotheses. As February 2020 preceded the outbreak of the COVID-19 pandemic, the data collected during this period represents a normal or controlled condition without perturbation by any crisis or significant impact. Human mobility networks in 11 metropolitan counties in the United States were examined: DeKalb of Georgia; DuPage of Illinois; Suffolk of Massachusetts; Wayne of Michigan; Bronx and Richmond (known as Staten Island) of New York; Clackamas and Washington of Oregon; Collin, Denton, and Kaufman of Texas.

\subsubsection{Spectus data}
The first data source is the Global Positioning System (GPS) location dataset obtained by Spectus from smartphone devices. Spectus amasses large-scale anonymous location data from almost 70 million U.S. mobile devices when users download one of its partner apps and opt in to the app's location services through a General Data Protection Regulation (GDPR) and California Consumer Privacy Act (CCPA) compliant framework. Users could choose to opt out of the app's location services at any time. Spectus partners with more than 220 mobile apps that include the proprietary Spectus software development kits (SDK). By collecting data from about one in four U.S. smartphones, Spectus covers almost 20\% of the U.S. population. The data were collected through partner applications and relied on devices’ internal GPS hardware. GPS sensor log data have been previously used as a source of data in fusion frameworks to study human mobility and in travel mode detection since such location data have a high spatiotemporal resolution. In addition to de-identifying data, Spectus applies additional enhancement to preserve privacy, such as obfuscating home locations to the census block group level, and removing sensitive Points of Interest (POI) from the data set. The device-level data contain individuals’ recorded location information, including an anonymized individual ID, location, and corresponding time (in seconds). Spectus makes available its own data and processing tools, as well as location-based datasets from other providers, through a Data Cleanroom environment for privacy preserving analyses of multiple geospatial data assets.

\subsubsection{X-Mode data}
X-Mode is a location intelligence company that collects the locations of anonymized mobile phone devices whose owners opt in to share their location information. One distinctive feature of X-Mode is their custom SDK: XDK. It is battery-efficient and extremely accurate when it comes to locating a device (within 20 to 30 meters). X-Mode has been collecting the data from 50 million mobile phone devices globally through General Data Protection Regulation-compliant and California Consumer Privacy Act-compliant frameworks. X-Mode collected data from more than 30 million mobile phone users with 2 billion to 3 billion location data every day in the United States. The X-Mode dataset is one of the most comprehensive location-based datasets of anonymized mobile devices. All records, anonymized and de-identified, contain data that includes information about the mobile devices, their locations and timestamps, their heading direction and speed, points of interest at their locations, and information about data sources.

\subsubsection{Veraset data}
Veraset movement data is provided by Veraset, Inc. Similar to Spectus, Veraset sources from thousands of apps and SDKs to collect granular device location points and to avoid biased samples. This dataset, covering more than 10\% of the US population, contains anonymized device IDs, timestamps, and precise geographical coordinates of dwelling points. The dwelling points, also called stop points, in the anonymized mobile phone data shared by the data company are defined as the points where the devices spent at least 5 minutes.

Researchers widely use these three datasets. Hence, we selected them in this study to compare their network analysis results obtained in different cities and at different scales.

\subsection{Human mobility network analysis}
The collected data from Spectus, X-Mode, and Veraset are first processed into daily trip counts from origins and destinations. The original form of the data only recorded the anonymized device ID, coordinate and visit time. For each device, every location of visit is determined based on which census tract polygon it falls in. By the precedence relationship obtained from visit times, a device’s movement, or trajectory is determined. The daily trip counts between census tracts are derived from daily aggregation on a census-tract level. We then use these daily trips counts to construct human mobility networks for further analyses. 
Consider an undirected network represented as $G=(V,E,w)$, where $V$ is the set of nodes, $E$ is the set of edges that connects each of the nodes, and w is the weight assigned to each of the edges. In this study, every node is the centroid of a census tract, and edges will be established if people are traveling from one census tract to another. w is the trip count between origin and destination. A total of 29 daily human mobility networks are generated for each county. These networks form the basis for macroscopic and substructure level analyses. Table 1 shows the sizes of the mobility network created from each data source for the 11 chosen counties. We can see that all the three data sources capture the same number of nodes while constructing mobility networks, but the number of edges (connections) captured might be significantly different, especially when the number of edges is large. Usually, X-Mode captures the most edges, while Veraset captures the fewest.

\begin{table}[]
\centering
\caption{Size of mobility networks created from Spectus, X-Mode, and Veraset for the 11 counties}
\begin{tabular}{llllllll}
\hline
FIPS &
  County &
  \begin{tabular}[c]{@{}l@{}}Number \\ of nodes \\ (Spectus)\end{tabular} &
  \begin{tabular}[c]{@{}l@{}}Number \\ of nodes \\ (X-Mode)\end{tabular} &
  \begin{tabular}[c]{@{}l@{}}Number \\ of nodes \\ (Veraset)\end{tabular} &
  \begin{tabular}[c]{@{}l@{}}Number \\ of edges \\ (Spectus)\end{tabular} &
  \begin{tabular}[c]{@{}l@{}}Number \\ of edges \\ (X-Mode)\end{tabular} &
  \begin{tabular}[c]{@{}l@{}}Number \\ of edges \\ (Veraset)\end{tabular} \\ \hline
13089 & DeKalb     & 144.90 & 144.86 & 144.97 & 5063.31  & 5493.21  & 4354.45  \\
17043 & DuPage     & 216.00 & 216.00 & 216.00 & 9140.76  & 9985.66  & 6675.76  \\
25025 & Suffolk    & 203.41 & 202.79 & 203.79 & 5752.41  & 7356.55  & 4487.48  \\
26163 & Wayne      & 609.10 & 609.00 & 609.10 & 30327.72 & 29563.14 & 20007.48 \\
36005 & Bronx      & 339.00 & 338.97 & 339.00 & 13535.69 & 15497.03 & 6959.59  \\
36085 & Richmond   & 109.00 & 108.97 & 109.00 & 4057.86  & 4188.79  & 2180.86  \\
41005 & Clackamas  & 80.00  & 80.00  & 80.00  & 1863.00  & 2064.86  & 1481.90  \\
41067 & Washington & 104.00 & 104.00 & 104.00 & 3397.48  & 3803.10  & 2706.86  \\
48085 & Collin     & 152.00 & 152.00 & 152.00 & 7404.03  & 8437.59  & 5549.48  \\
48121 & Denton     & 137.00 & 137.00 & 137.00 & 5244.17  & 5871.76  & 3932.93  \\
48257 & Kaufman    & 18.00  & 18.00  & 18.00  & 150.86   & 170.00   & 139.03   \\ \hline
\end{tabular}
\end{table}

\subsection{Macroscopic-level analyses}
The macroscopic level mobility features—average degree, average shortest path length, clustering coefficient and assortativity coefficient calculated from daily human mobility networks—can all be represented as times series from February 1 to February 29. By observing the similarities of their time-series patterns, we can examine similarities among the results obtained from different datasets. The Euclidean distance metric and mean absolute percentage error (MAPE) are based on calculating the difference between two time series point-by-point. The value of Euclidean distance can range from 0 (identical time series) to infinity. Euclidean distance depends not only on the similarity between two time series but also on their length or number of points compared. MAPE performs normalization by taking the mean of the point-by-point distance divided by the base value. This definition is not symmetric, which means MAPE differs when different time series are chosen to be the base. This limitation can be alleviated by dividing each point-by-point distance by the average value of both time series instead of dividing by the base directly. Another problem with the point-by-point comparisons of absolute values is that the offset or scaled curves will simply be identified as dissimilar because no knowledge from neighboring points is considered \citep{deza_encyclopedia_2012}. The Pearson’s correlation coefficient also does point-by-point comparisons and measures the linear correlation between datasets. Compared with the point-by-point comparisons between different series, dynamic time warping (DTW) is an optimized way of trying a subset of all reasonable time mappings between two time series and choosing the best match \citep{muller_dynamic_2007}. Hence, in this study, we used DTW for comparison of results among models constructed from different datasets.

\subsection{Substructure-level analyses}
Motifs are statistically overrepresented substructures (subgraphs) in a network and have been recognized as simple building blocks of complex networks \citep{artzy-randrup_comment_2004,schneider_unravelling_2013}. A motif $G_m$ is defined as a recurrent multi-node-induced subgraph in network $G$. In this study, we consider motifs constructed with four-node pairs. These four-node pairs are classified into seven motif types (Figure 1) throughout our substructure-level analyses. Motif types 1 and 2 represent densely connected subgraphs where almost all nodes are connected. They represent movement between areas that exhibit the most interconnected movement patterns. Motif type 4 has a three-node cycle and an open edge. Motif types 3 and 5 may represent general commute patterns for work or other lifestyle patterns. Motif type 6 represents a hub and a spoke structure. Motif type 0 is a symbolic motif type that we use to identify four-node pairs where at least one node is disconnected from its subgraph and does not fall into any of the six types.

\begin{figure}
	\centering
    \includegraphics[width=0.8\linewidth]{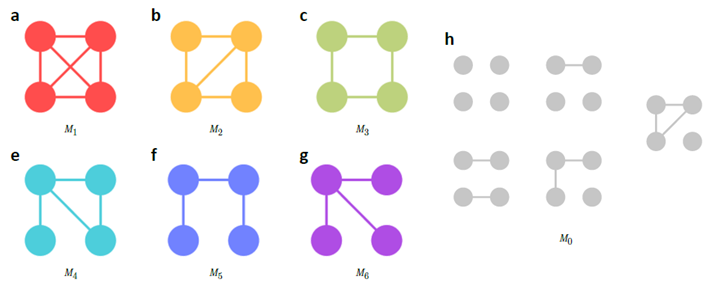}
    \caption{Seven types of four-node pair motifs examined in this study}
	\label{fig:fig1}
\end{figure}

\subsubsection{Motif distribution}
Motif count for each human mobility network is recorded for all counties in February 2020. Previous research working with a different data source stated the difficulty of counting every motif presence because of the size and complexity of networks. They found that sampling 100,000 four-node pairs may accelerate their computation without harming the ability to represent the entire network \citep{rajput_latent_2022}. However, since the purpose of this study is to compare the same features between different datasets, we may not be able to conclude a single set of suitable sampling attempts that can produce stable motif count results for each dataset in each county. Thus, we will still record every motif presence instead of sampling the four-node pairs. Motif distribution is defined as the number of occurrences of motif types divided by the total number of node pairs. To reduce the noise in the data, we also calculated a seven-day moving average for each day; the results for weekdays and weekends are viewed separately. The key features we are interested in are the proportion of motif types and their stability.

\subsubsection{Motif attributes}
To understand the extent to which the intrinsic characteristics of these motifs change over time and how they shape the characteristics of human mobility networks, we examined the change in average travel volume and the average distance metric for each motif type \citep{gonzalez_unraveling_2013}. Average motif distance represents the distance range that a particular motif type connects to on average. To compute this metric, we first geocode all the nodes in the network based on the location of the nodes that represent the centroids of census tracts. Then, we calculate the distance represented by each of the links in the motif subgraph and calculate its average. For all the motifs of the same type, from the distribution of these average distance values, we report the median value. Average travel volume for a motif is calculated by adding all the link weights for a motif subgraph. This metric represents the average travel volume that a motif is contributing to the entire network. First, we collect all four-node pairs and split the four-node pairs into six motif types based on their corresponding motif. For node pairs in each of these motif types, we get the motif subgraph and calculate the average of the edge weights (travel volume). We report the median of these average values in each motif type. Absolute values of computed measures are plotted for all 29 days across all examined metro area. 

\subsection{Microscopic-level analyses}
At the microscopic level, we investigate individual device movements. Each trip of every device is recorded and then aggregated to census-tract level. For each census tract, we filter the trips that start or end in that census tract. Then, we calculate network characteristics for each census tract based on those filtered trips. The network characteristics we examined in this step are travel volume, travel distance, travel time, and radius of gyration \citep{gonzalez_understanding_2008,hoteit_estimating_2014}. These characteristics are calculated for each census tract daily in different counties. We applied the cosine similarity method to quantify the similarity of these characteristics between different data sources. Cosine similarity is a measure of similarity between two sequences of numbers and captures the cosine of the angle between two n-dimensional vectors in an n-dimensional space. It is the dot product of the two vectors divided by the product of the two vectors' lengths (or magnitudes). Values range between -1 and 1, where -1 is perfectly dissimilar and 1 is perfectly similar.

\section{Results}
\label{sec:Results}
\subsection{Macroscopic-level analyses}
In this section, we investigate global characteristics (average degree, average clustering coefficient, average shortest path length, and assortativity coefficient) of human mobility networks constructed from Spectus, X-Mode, and Veraset data. For each characteristic, we applied DTW to compare the similarity between results calculated from each data source. Table 2 shows which two data sources produce more comparable results for each characteristic in counties. The notations SV (Spectus–Veraset), XV (X-Mode–Veraset), SX (Spectus–X-Mode) represent which two data sources usually yield more comparable results based on DTW analysis. We can observe that for average degree and average shortest path length, Spectus and X-Mode data yield comparable results in most counties, while Spectus and Veraset data produce comparable results on assortativity coefficient in most counties. However, the absolute value of assortativity coefficient is usually very small compared to the other characteristics and oscillates around 0, so the result may be somehow biased. For clustering coefficient, on the other hand, we cannot find a dominant data source pair that usually has results more similar to each other. These results suggest that Spectus and X-Mode could produce comparable results for most global human mobility network measures, while the network measures obtained from Veraset could be different and incomparable.

\begin{table}[]
\centering
\caption{Similarity between data sources regarding macroscopic features}
\label{tab:my-table}
\begin{tabular}{llllll}
\hline
FIPS &
  County &
  \begin{tabular}[c]{@{}l@{}}Average \\ degree\end{tabular} &
  \begin{tabular}[c]{@{}l@{}}Average \\ clustering \\ coefficient\end{tabular} &
  \begin{tabular}[c]{@{}l@{}}Average \\ shortest path \\ length\end{tabular} &
  \begin{tabular}[c]{@{}l@{}}Assortativity \\ coefficient\end{tabular} \\ \hline
13089 & DeKalb     & SX & SX & SX & SV \\
17043 & DuPage     & SX & SV & SX & SV \\
25025 & Suffolk    & SX & SX & SX & SX \\
26163 & Wayne      & SX & SV & SX & SV \\
36005 & Bronx      & SX & SX & SX & SV \\
36085 & Richmond   & XV & XV & XV & SV \\
41005 & Clackamas  & SX & SX & SX & SV \\
41067 & Washington & SX & XV & SX & SV \\
48085 & Collin     & SX & SX & SX & SV \\
48121 & Denton     & SX & SX & SX & SX \\
48257 & Kaufman    & SV & SV & SV & SX \\ \hline
\end{tabular}
\end{table}

\subsection{Substructure level analyses}
\subsubsection{Motif distribution}
Figure 2 shows the motif distribution of daily human mobility networks acquired from Spectus, X-Mode, and Veraset for DuPage County, Illinois, on February 3 and February 8, 2020, which represent weekdays and weekends, respectively. The distribution patterns on weekdays and weekends are the same but differ between the three datasets. The major motif types are type 4 and 5 for Spectus and Veraset. For X-Mode, the occurrence of type 1 motif, that accounts for more than 50\% of motifs, is significantly more frequent than other types. Spectus and Veraset data give similar results, while the X-Mode results appear to be different. Similar patterns can be found in most of the counties; however, Figure 3 shows another example in Richmond, New York, in which we no longer see the resemblance between either of the two data sources. In the X-Mode motif distribution for all the counties, the majority of motif types are type 1 followed by types 2 and 3, and the rest accounted for only a very small proportion. This implies that the substructure of the human mobility networks constructed from X-Mode data are more densely connected. Figures 4 to 6 show the percentage change of motif distribution in DuPage County for three data sources. We can see that the percentages of more frequent motif types are more stable, while motif types that account for smaller proportions fluctuate more for all three data sources.

\begin{figure}
	\centering
    \includegraphics[width=0.95\linewidth]{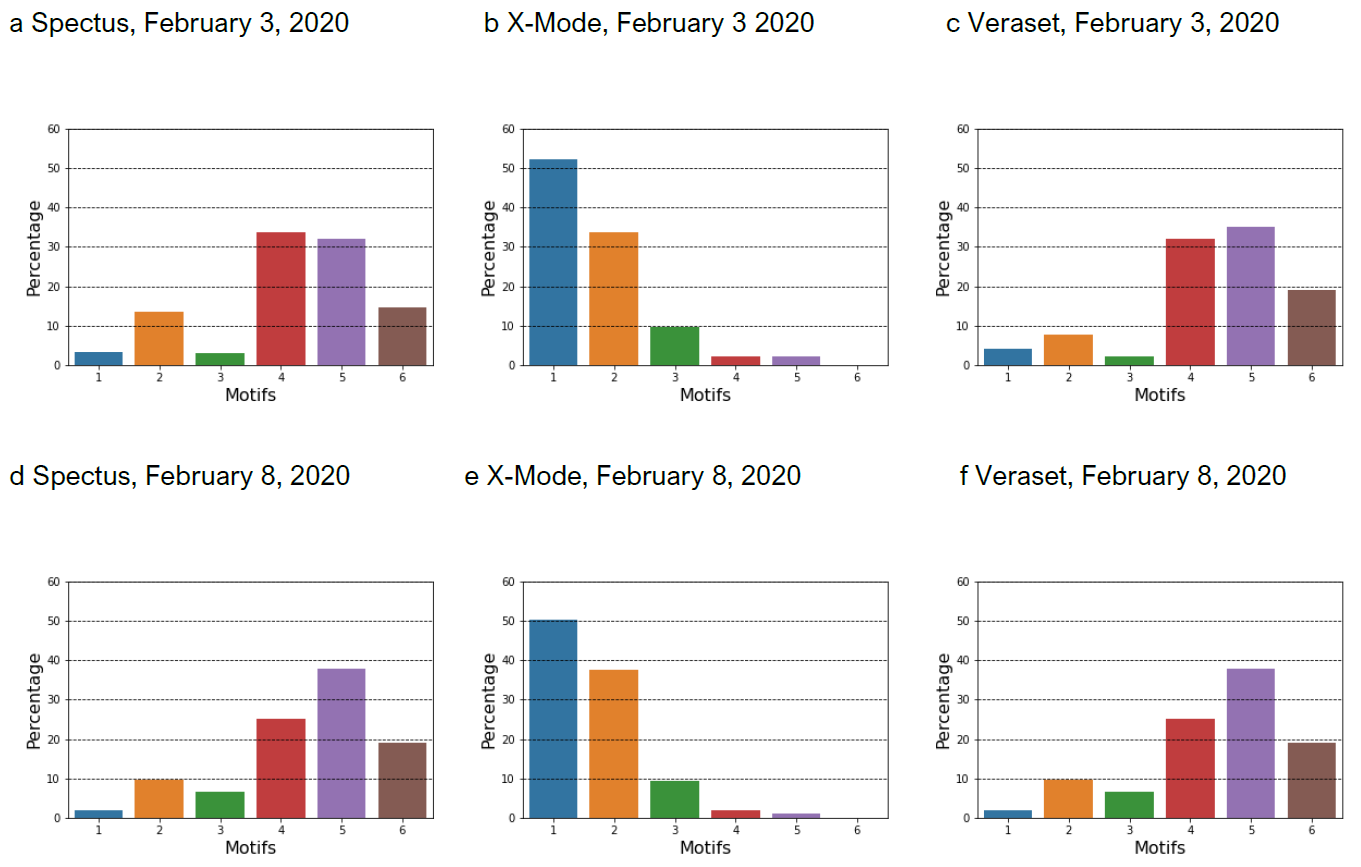}
    \caption{Distribution of motif types in terms of relative occurrence from daily human mobility networks constructed from datasets for DuPage County. (a) Spectus for February 3, 2020; (b) X-Mode for February 3, 2020; (c) Veraset for February 3, 2020; (d) Spectus for February 8, 2020; (e) X-Mode for February 8, 2020; (f) Veraset for February 8, 2020}
	\label{fig:fig2}
\end{figure}

\begin{figure}
	\centering
    \includegraphics[width=0.8\linewidth]{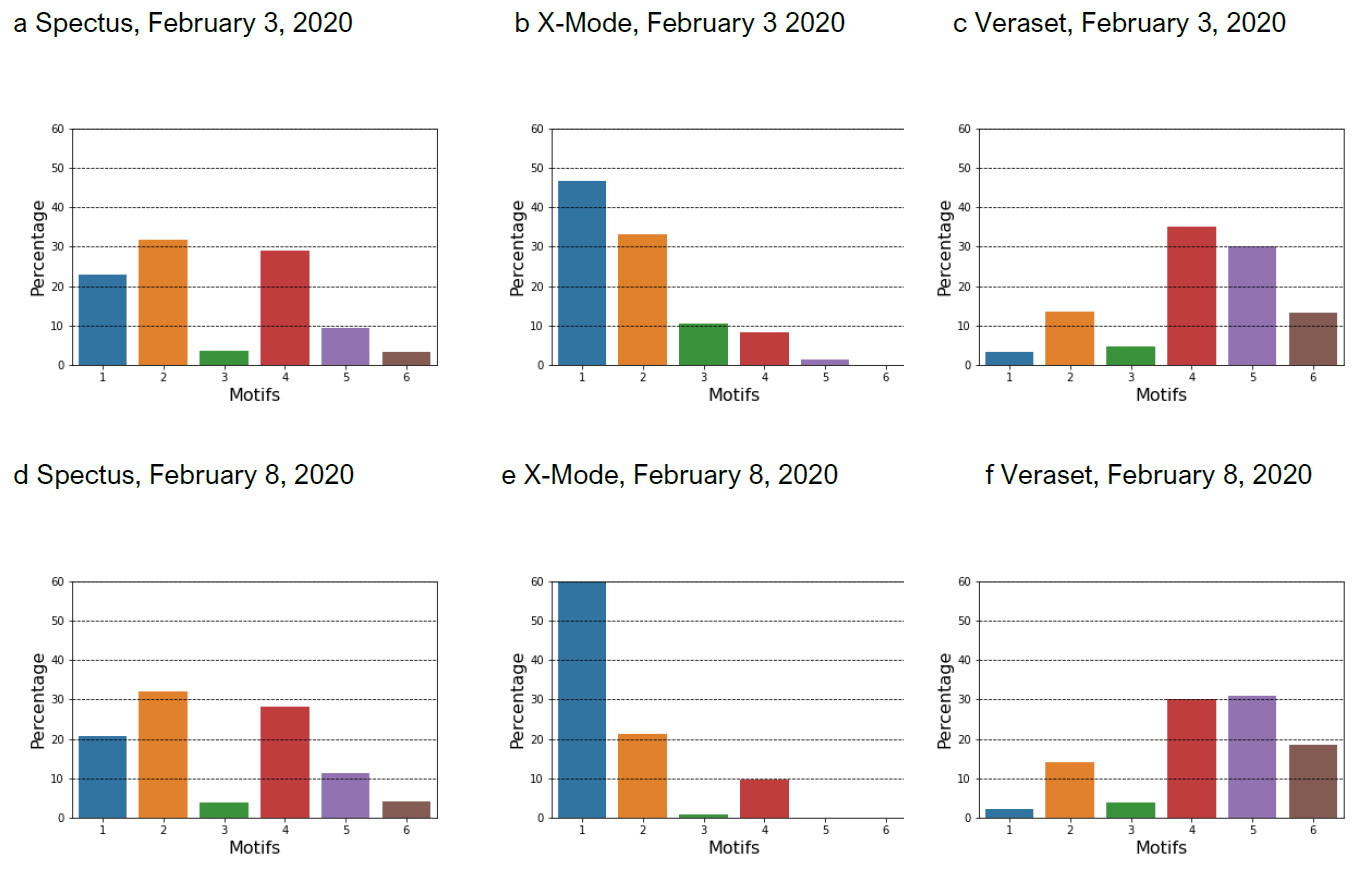}
    \caption{Distribution of motif types in terms of relative occurrence from daily human mobility network constructed from datasets for Richmond County. (a) Spectus for February 3, 2020; (b) X-Mode for February 3, 2020; (c) Veraset for February 3, 2020; (d) Spectus for February 8, 2020; (e) X-Mode for February 8, 2020; (f) Veraset for February 8, 2020}
	\label{fig:fig3}
\end{figure}

\begin{figure}
	\centering
    \includegraphics[width=0.8\linewidth]{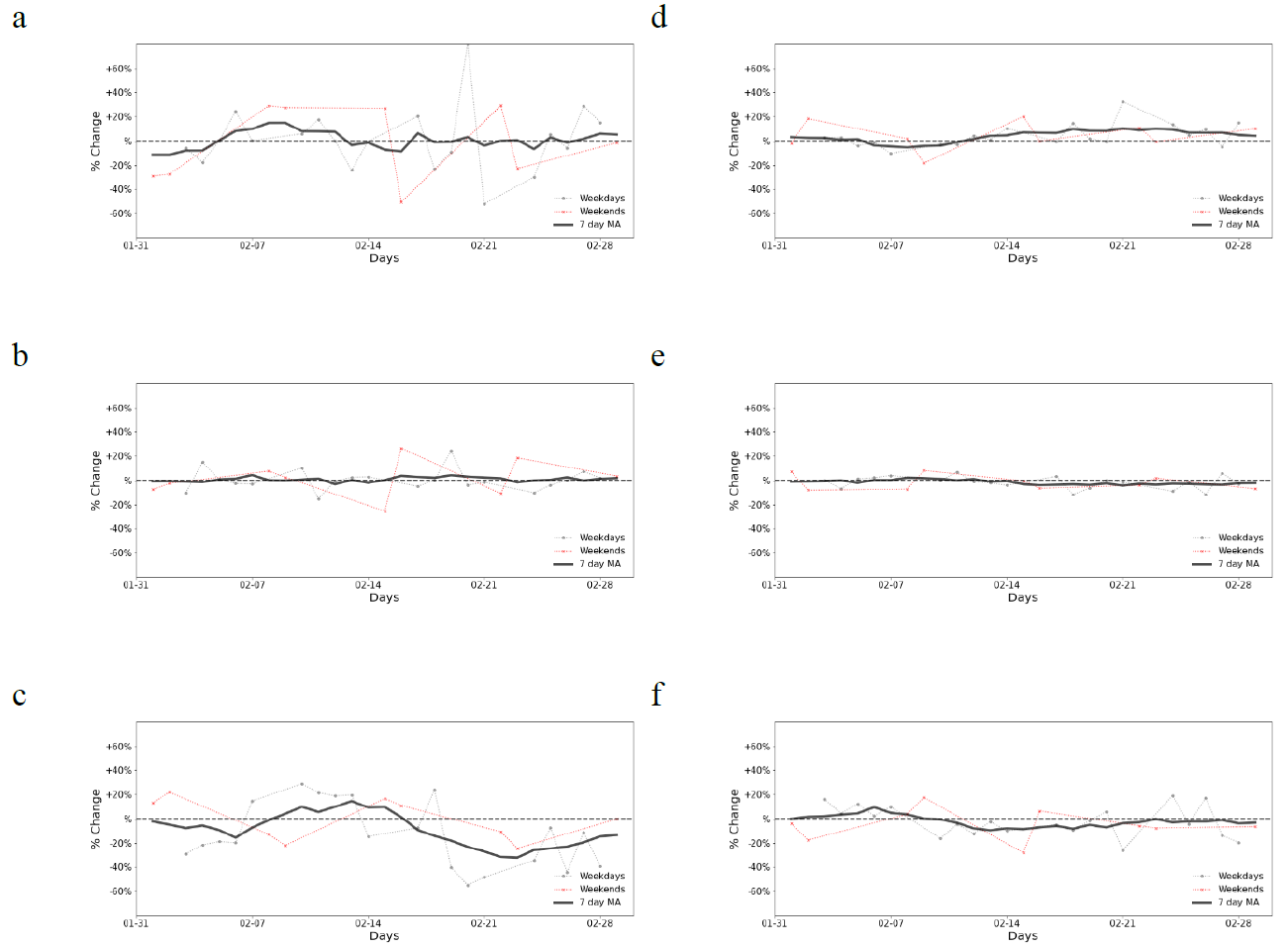}
    \caption{Spectus motif distribution for DuPage County. (a)–(f) show the change in motif distribution for motifs 1–6, respectively.}
	\label{fig:fig4}
\end{figure}

\begin{figure}
	\centering
    \includegraphics[width=0.8\linewidth]{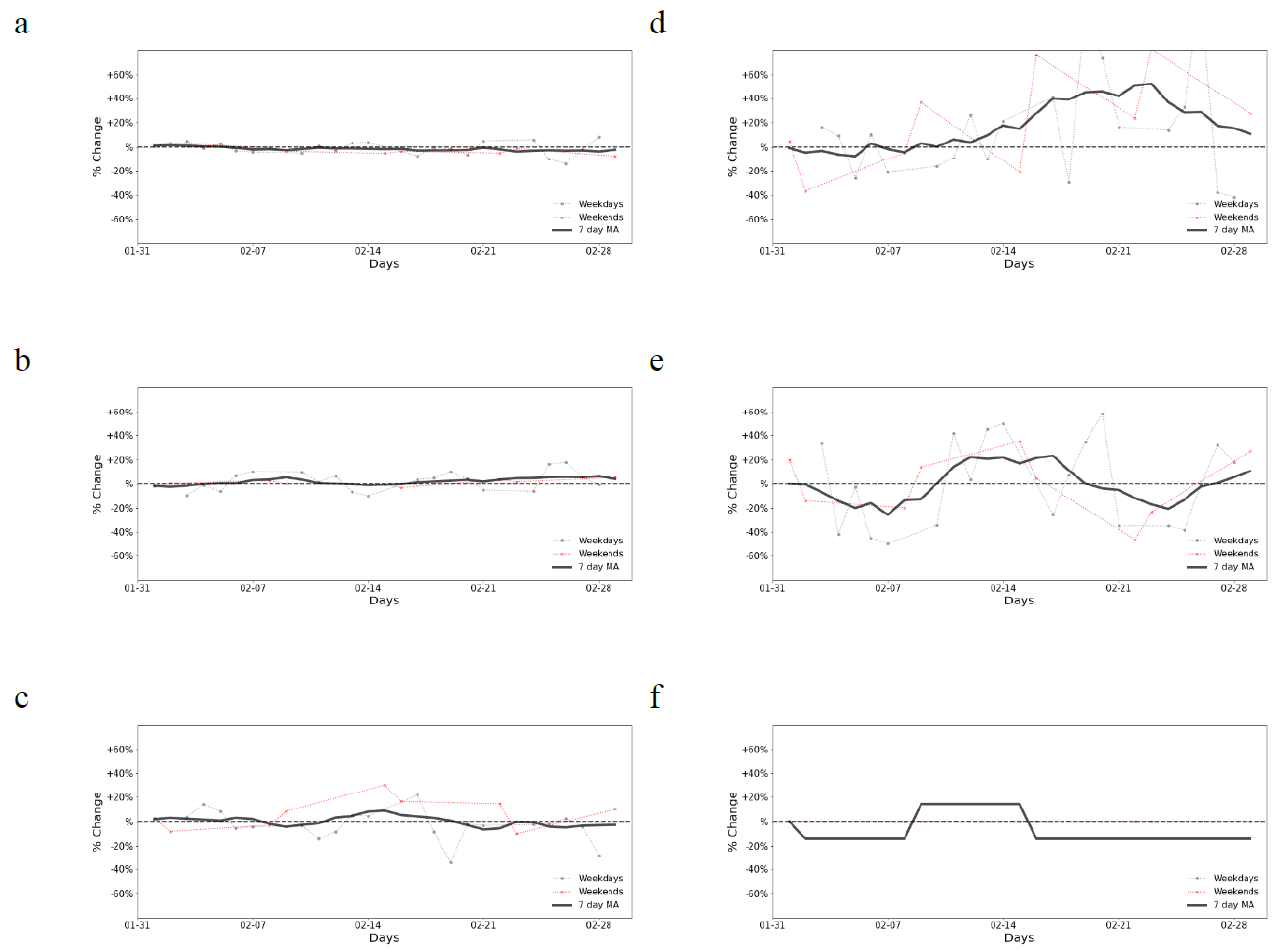}
    \caption{X-Mode motif distribution for DuPage County. (a)–(f) show the change in motif distribution for motifs 1–6, respectively.}
	\label{fig:fig5}
\end{figure}

\begin{figure}
	\centering
    \includegraphics[width=0.8\linewidth]{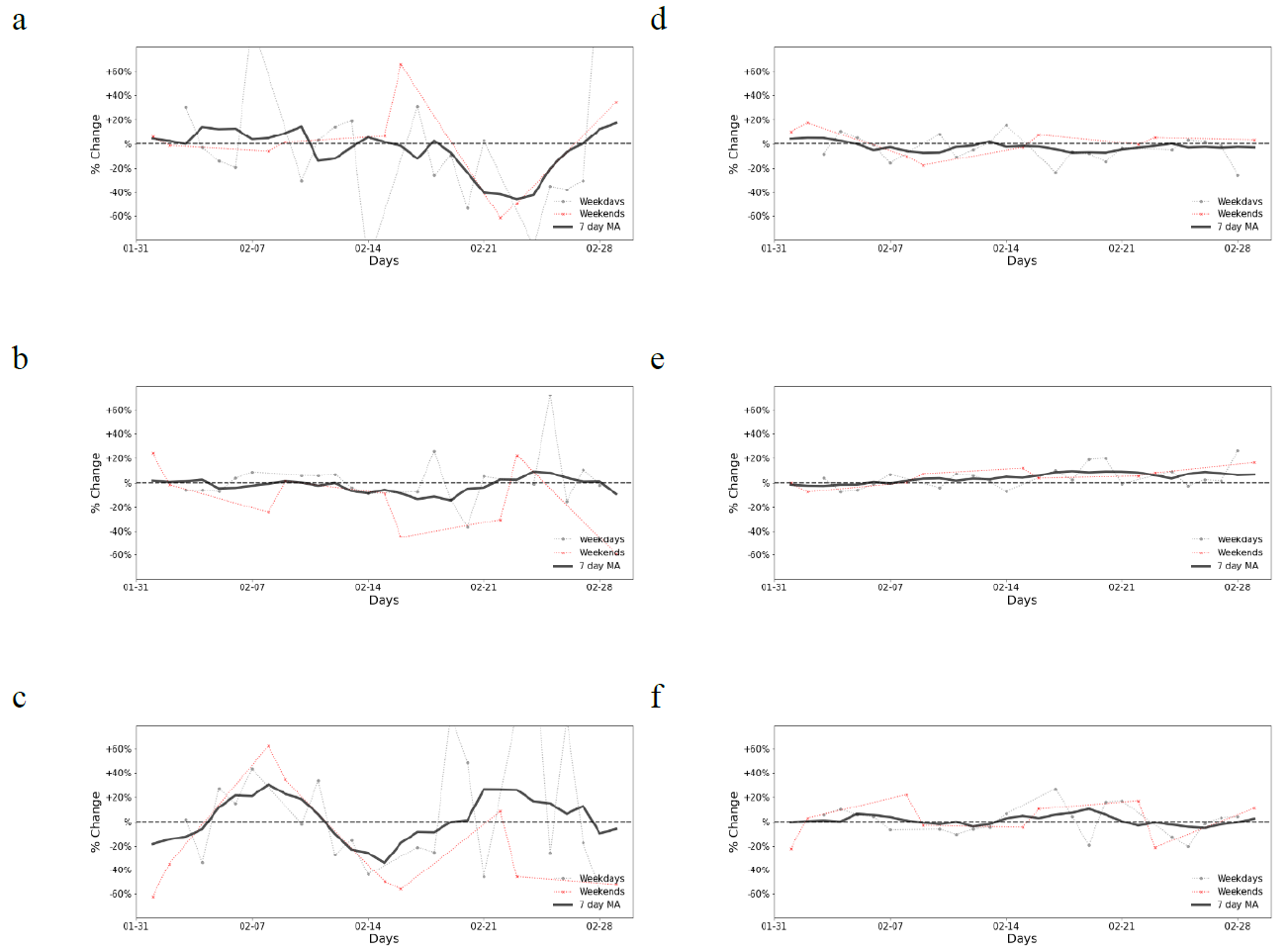}
    \caption{Veraset motif distribution change of different motif types for DuPage County. (a)-(f) show the change in motif distribution for motifs 1-6 respectively.}
	\label{fig:fig6}
\end{figure}

\subsubsection{Motif attributes}
Next, we evaluate the sensitivity of motif properties, such as average travel distance and average travel volume change, with time from different data sources across counties. Figure 7 shows the absolute values for the median of average motif distance. From the results generated from Spectus and Veraset (Figures 7 a and c), motif type 6 and type 3 connect the farthest places, followed by motif type 4 and type 5. Motif type 1 and type 2 connect the nearest places. The results for X-Mode (Figure 7 b), however, shows the opposite. During weekends, the distance reduces for all motif types, indicating shorter trips, while motif distribution remains the same. We cannot find a strong association between the temporal stability of motif types and their travel distances. Contrasting but intuitive trends can be found in travel volume (Figure 8) for Spectus and Veraset (Figures 8 a and c) compared to travel distance. Motif type 1 corresponds to the highest travel volume; motifs 6, 5, and 3 have the lowest travel volume. Figures for X-Mode (Figure 8 b), again, show the opposite trend.

\begin{figure}
	\centering
    \includegraphics[width=0.65\linewidth]{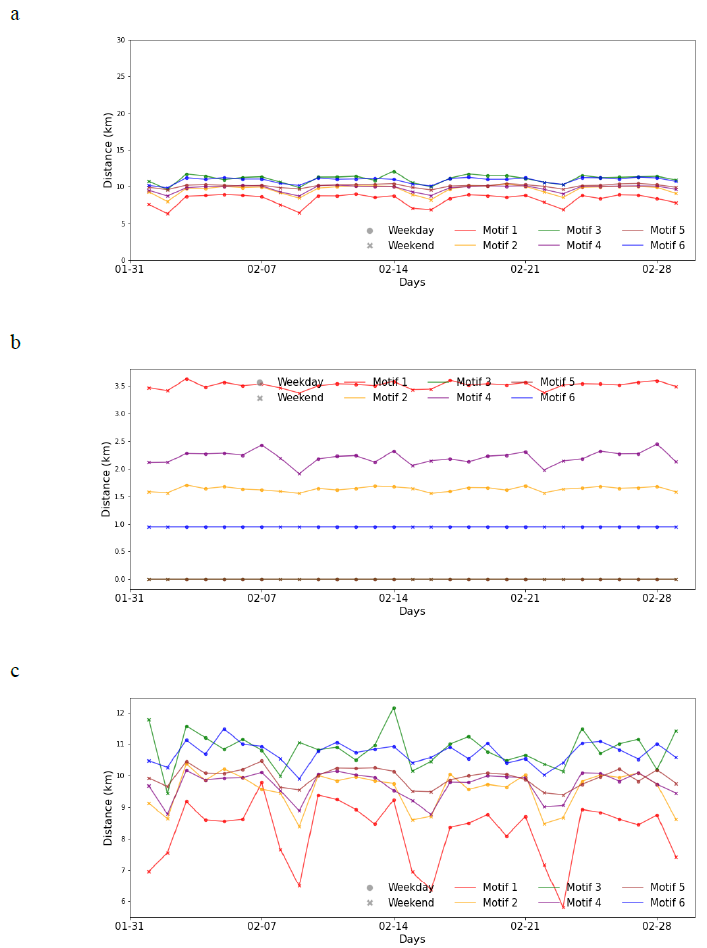}
    \caption{Absolute values of median of the average motif distance for DuPage County. The figure represents the median average distance that each of the motifs connects spatially. (a) Spectus; (b) X-Mode; (c) Veraset.}
	\label{fig:fig7}
\end{figure}

\begin{figure}
	\centering
    \includegraphics[width=0.65\linewidth]{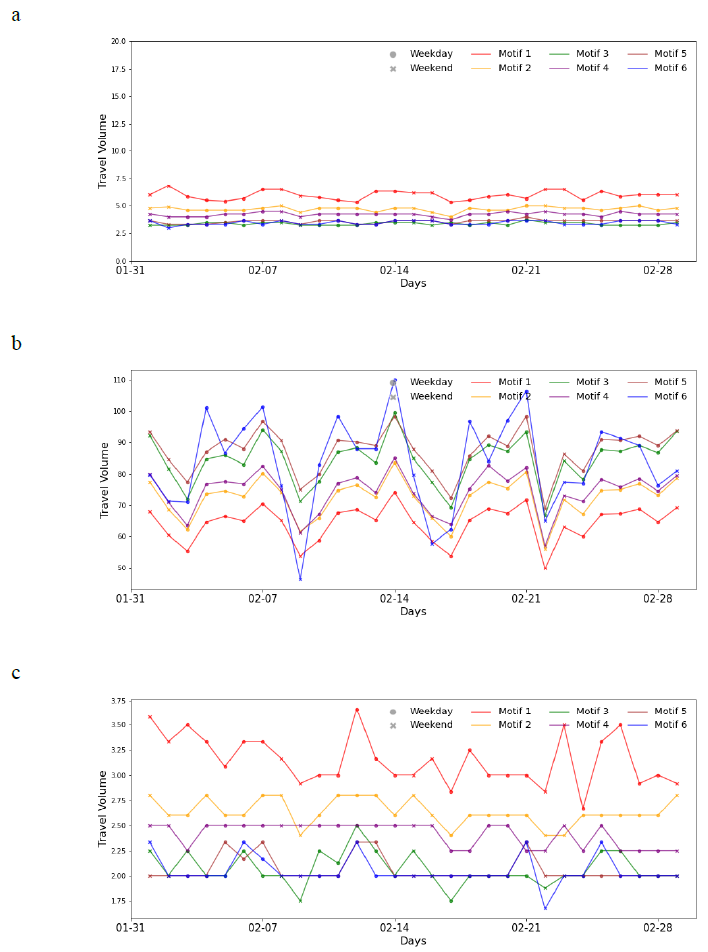}
    \caption{Absolute values of the median of the average motif travel volume for DuPage County. The figure represents the median average travel volume that each of the motifs caters to. Denser motifs that have one or more three-node cycles tend to have higher travel volumes. (a) Spectus; (b) X-Mode; (c) Veraset.}
	\label{fig:fig8}
\end{figure}

\subsection{Microscopic-level analyses}
After metrics for the microscopic characteristics (average number of trips, average trip distance, average travel time, and average radius of gyration) are calculated, we rank the census tracts in each county based on the values of each metric. Then we perform cosine similarity analysis daily between each data source pair. Table 3 shows the results of cosine similarity analysis between data sources across counties. The notations SV (Spectus–Veraset), XV (X-Mode–Veraset), and SX’ (Spectus–X-Mode) represent the pair of data sources usually yielding more similar results based on cosine similarity. Spectus and Veraset give similar results on average number of trips across all counties. X-Mode and Veraset give similar results on average travel time across all counties; however, the similarity of average number of trips is not as close as that of Spectus and Veraset. For the other two metrics, X-Mode and Veraset are more similar on average trip distance while Spectus and Veraset are more similar on average radius of gyration. The similarity between Spectus and Veraset is usually greater while the Spectus–X-Mode similarity and X-Mode–Veraset similarity do not show strong evidence of close similarity. Except for average number of trips and average travel time, there does not exist any general association between the similarity of data sources and microscopic characteristics.

\begin{table}[]
\centering
\caption{Similarity between data sources regarding microscopic features}
\label{tab:my-table}
\begin{tabular}{llllll}
\hline
FIPS &
  County &
  \begin{tabular}[c]{@{}l@{}}Average \\ Number \\ of Trips\end{tabular} &
  \begin{tabular}[c]{@{}l@{}}Average \\ Travel \\ Distance\end{tabular} &
  \begin{tabular}[c]{@{}l@{}}Average \\ Travel \\ Time\end{tabular} &
  \begin{tabular}[c]{@{}l@{}}Average \\ Radius of \\ Gyration\end{tabular} \\ \hline
13089 & DeKalb     & SV & SX & XV & SX \\
17043 & DuPage     & SV & XV & XV & SX \\
25025 & Suffolk    & SV & SV & XV & SV \\
26163 & Wayne      & SV & XV & XV & XV \\
36005 & Bronx      & SV & SV & SX & SV \\
36085 & Richmond   & SV & XV & XV & SX \\
41005 & Clackamas  & SV & XV & XV & SV \\
41067 & Washington & SV & XV & XV & SX \\
48085 & Collin     & SV & XV & XV & SX \\
48121 & Denton     & SV & XV & XV & SV \\
48257 & Kaufman    & SV & SX & XV & SX \\ \hline
\end{tabular}
\end{table}

\section{Discussion and concluding remarks}
\label{sec:Discussion and concluding remarks}
Analysis of human mobility networks has proven useful in the fields of science and engineering, as well as found practical application in epidemic mitigation and urban planning. Studies using different datasets to construct human mobility networks may arrive at different results. Hence, the models and findings based solely on a single dataset may not yield generalizable insights. Thus, it is essential to examine whether human mobility network models constructed from different datasets yield similar results in order to compare and generalize results and insights of disparate studies. In this study, we utilized high-resolution aggregated location-based data sources from three major providers in the United States to construct daily human mobility networks in 11 counties during February 2020, which represents a non-perturbed condition with no significant natural hazards or crises. Spectus, X-Mode, and Veraset are three widely used data sources in studying human mobility-related research topics. Based on Spectus, X-Mode and Veraset data, we acquired global network characteristics, properties of substructures, and microscopic characteristics. We hypothesized that all data sources may show strong similarities on higher level of network analyses since human movements are aggregated, and the structure of human mobility networks should be captured alike by different datasets. The analysis of global characteristics suggests, however, that even for the network-level measures, different data sources do not yield the same results across different counties. Spectus and X-Mode usually yield similar results on average degree and average shortest path length, but we cannot conclude which two data sources give us more similar results on average clustering coefficient and assortativity coefficient. The analysis suggests that, in the main, the findings concluded from Spectus and X-Mode data related to certain global network characteristics may be comparable and transferrable. 

For the substructure-level analysis, we investigated motif distribution and the properties of each motif type. Patterns of motif distribution remain the same in a county on weekdays and weekends but the patterns differ among the three data sources. In most counties, Spectus and Veraset data give somewhat similar results of motif distributions while X-Mode results appear to be different; nevertheless, the three data sources give entirely different distributions. From the distributions, we can see that the complexity or connection density of a motif type does not seem to have a clear association with the occurrence frequency. Another observation is that the percentages of major motif types are more stable, while motif types that account for smaller proportions fluctuate more for all the three data sources. Overall, motif distribution for X-Mode is usually different from the others. For the motif properties, again, Spectus and Veraset give similar results. Motif types with the longest distance in Spectus and Veraset have the lowest travel volume. Motifs in the X-Mode data, on the other hand, have a much shorter travel distance and larger volumes compared to Spectus and Veraset. Based on these observations, we can conclude that comparison of the results of human mobility network motifs built from different datasets does not yield similar patterns and insights.

For the microscopic level analysis, we found that Spectus and Veraset give similar results on average number of trips, and X-Mode and Veraset give us similar results on average travel distance across most counties and on average travel time across all counties. The calculation of average travel time is calculated from the difference between the stop time of origin and destination, then subtracting the dwelling time, which might be the largest error source. Thus, we conclude that the three datasets do not produce similar results for microscopic characteristics. The results show that, first, there is no clear evidence that all the three datasets produce similar results across the analysis scales. At each level of analysis, we found dataset pairs that produce similar results on different features or types of analyses. This findings highlight the importance of having more than two datasets for evaluating whether certain patterns and insights do really exist in human mobility networks. Second, the differences between results acquired from datasets using different technologies (SDK or XDK) are distinguishable at the substructure level but not at coarser or finer level. We know that X-Mode technology is different from Spectus and Veraset, and that the resulting data size and types of trips (regarding travel distance) are also different. This may be the reason X-Mode yields different results for the motif analysis. This finding highlights the need for representative and comparable human mobility datasets to be used to advance theories of human movement and urban dynamics across different fields of science and engineering.

This study and results provide important evidence on whether analyses results based on high-resolution location-based data and network analysis are comparable and transferrable. The significance of the results lies at the need for ground-truthed human movement data to build generalizable and transferable theories of human mobility and urban dynamics across different scales. Results derived from a specific dataset cannot be generalized in most cases. A single dataset can only represent part of the entire human mobility network which might be biased when looking for general patterns due to their data collection technology, penetration rate and targeted users. Thus, it is essential that studies based on these location-based datasets should the results among the comparable datasets to understand if there are fundamentals differences in the data characteristics itself. By doing so, we may be able to conclude which dataset is more representative on certain types of analyses or areas that produce comparable results. While the existing datasets provided by various providers such as Spectus provide representative data of human mobility to inform research and decisions, there is a need for datasets that can be ground-truthed. They can serve as benchmarks that help examine data validity before expanding spatial-temporal mobility pattern analyses to other areas under different contexts. For example, the location-based data could be used to create machine-learning and agent-based models that could simulate more representative patterns of human movements and trajectories to be used for data-driven discoveries around human mobility. As we keep collecting human mobility data, a much stronger human mobility pattern basis, as well as its change through time, can be obtained. With the increasing amount of human mobility data, better understanding on the characteristics of each dataset and systematic validation techniques, human mobility-driven insights and theories can be more trustworthy, generalizable, and transferable.

\section*{Acknowledgement}
This material is based in part upon work supported by the National Science Foundation under Grant 2026814 and the Texas A\&M University X-Grant 699. The authors also would like to acknowledge the data support from Spectus, Inc, X-Mode, and Veraset. Any opinions, findings, conclusions, or recommendations expressed in this material are those of the authors and do not necessarily reflect the views of the National Science Foundation, Texas A\&M University, Microsoft Azure or Spectus.

\section*{Data availability}
All data were collected through a CCPA and GDPR compliant framework and utilized for research purposes. The data that support the findings of this study are available from Spectus, X-Mode, and Veraset, but restrictions apply to the availability of these data, which were used under license for the current study. The data can be accessed upon request submitted to the providers. The data was shared under a strict contracts through their academic collaborative program, in which they provide access to de-identified and privacy-enhanced mobility data for academic research. All researchers processed and analyzed the data under a non-disclosure agreement and were obligated not to share data further or to attempt to re-identify data.

\section*{Code availability}
The code that supports the findings of this study is available from the corresponding author upon request.

\bibliography{ref}  





\pagebreak

\end{document}